\documentclass[runningheads]{llncs}
\usepackage{graphicx}

\usepackage{amsfonts}

\usepackage{algorithm}
\usepackage{algpseudocode}
\usepackage{cases}
\usepackage{url}
\usepackage[colorlinks=true, allcolors=blue]{hyperref}

\begin{document}

\title{Efficient Neural Network Approximation \\
of Robust PCA for Automated Analysis\\
of Calcium Imaging Data}

\titlerunning{Efficient Neural Network Approximation of Robust PCA}

\author{Seungjae Han$^1$ \and Eun-Seo Cho$^1$ \and Inkyu Park$^2$ \and Kijung Shin$^{1,2}$ \and \\
Young-Gyu Yoon$^1$}

\authorrunning{S. Han et al.}

\institute{$^1$School of Electrical Engineering, KAIST\\
$^2$Graduate School of AI, KAIST\\
\email{ygyoon@kaist.ac.kr}}

\maketitle
\begin{abstract}
Calcium imaging is an essential tool to study the activity of neuronal populations.
However, the high level of background fluorescence in images hinders the accurate
identification of neurons and the extraction of neuronal activities.
While robust principal component analysis (RPCA) is a promising method that can decompose the foreground and background in such images,
its computational complexity and memory requirement are prohibitively high to process large-scale calcium imaging data.
Here, we propose BEAR, a simple bilinear neural network for the efficient approximation of RPCA
which achieves an order of magnitude speed improvement with GPU acceleration compared to the conventional RPCA algorithms.
In addition, we show that BEAR can perform foreground-background separation of calcium imaging data as large as tens of gigabytes.
We also demonstrate that two BEARs can be cascaded to perform simultaneous RPCA and non-negative matrix factorization for the automated extraction of spatial and temporal footprints from calcium imaging data. The source code  used in the paper is available at \url{https://github.com/NICALab/BEAR}.

\end{abstract}

\section{Introduction} 
Recent advances in calcium imaging techniques have enabled imaging population neuronal activity across a large volume~\cite{stevenson2011advances}. 
State-of-the-art functional imaging methods can generate more than a gigabyte of data per second~\cite{bouchard2015swept,cong2017rapid,voleti2019real,yoon2020sparse} that is prohibitively large to be analyzed by humans, which necessitates the development of automated analysis algorithms. 
Furthermore, calcium imaging data suffers from a high level of background fluorescence, which stems from the intrinsic property of calcium indicator molecules.
Neurons with a calcium indicator have a certain baseline fluorescence level at rest (i.e., no activity),
which becomes brighter by 20 to 40\% when there is a single spike. 
Because most of the neurons are silent at a given time, the images consist of mostly background and a very small portion of foreground,
which makes it challenging to identify neurons in calcium imaging data.
Therefore, it is desirable to first separate the foreground (i.e., activity) from the backgrounds~\cite{li2017fast,peng2014shading}
before applying downstream analysis algorithms, such as image-based neuron detection algorithms~\cite{dong2020towards,kirschbaum2020disco},
NMF-based signal extraction methods~\cite{pnevmatikakis2016simultaneous,nejatbakhsh2020demixing}, and even volume reconstruction~\cite{yoon2020sparse}.

Robust principal component analysis (RPCA)~\cite{candes2011robust} is a promising method 
for the separation of the foreground in calcium imaging data. 
Neuronal activities can be separated from the background fluorescence
by exploiting that the background is nearly stationary, while the activity is spatiotemporally sparse~\cite{yoon2020sparse}. 
The separation can be formulated as the following optimization problem:
\begin{equation}
\label{equation_1}
\min_{L, S}(rank(L)+\lambda ||S||_{0}) \textrm{ subject to } Y=L+S,
\end{equation}
where $Y$, L, $S$, $||S||_0$, and $\lambda$ 
are a data matrix, a low-rank matrix, a sparse matrix, 
the $L_0$ norm of $S$, and a hyperparameter, respectively.
While this problem is known to be computationally intractable,
the exact solution can be obtained under weak assumptions through the surrogate optimization as follows ~\cite{candes2011robust,chandrasekaran2011rank}: 
\begin{equation}
\label{equation_2}
\min_{L, S}(||L||_{*}+\lambda ||S||_{1}) \textrm{ subject to } Y=L+S,
\end{equation}
where $||L||_{*}$ is the nuclear norm of $L$, and $||S||_1$ is the $L_1$ norm of $S$.

However, conventional RPCA based on the principal component pursuit method (PCP)~\cite{candes2011robust} for solving the optimization problem 
has limited capabilities when it comes to handling extremely large data because
it involves singular value decomposition (SVD) of the entire data matrix, which is computationally expensive.
In addition, solving the exact optimization problem requires storage of the entire data in the main memory,
which is not feasible for large-scale calcium imaging data sets.
Thus, many variants of RPCA have been developed to increase the speed ~\cite{hovhannisyan2017multilevel,zhou2011godec,zhou2013greedy} or to process large data~\cite{feng2013online,rahmani2017high,xiao2019online},
but it remains a challenge to simultaneously achieve
high speed and scalability.

Here, we introduce a Bilinear neural network for Efficient Approximation of RPCA (BEAR) 
which is a computationally efficient implementation of RPCA as a neural network.
BEAR has a number of advantages over conventional RPCA algorithms.
It (a) has significantly lower time complexity of $O(nmr)$ than conventional RPCA with time complexity of $O\left(nm^2+n^2m\right)$ for single iteration, 
where $n$ and $m$ are the sizes of the data along two dimensions and $r$ is the rank of $L$ which is typically a small integer,
(b) can decompose extremely large data,
and (c) can be naturally combined with other neural networks for end-to-end training. 

\section{Methods}
\subsection{BEAR as a Robust Principal Component Analysis} \label{2.1}

\begin{figure}[t!]
\includegraphics[width=\textwidth]{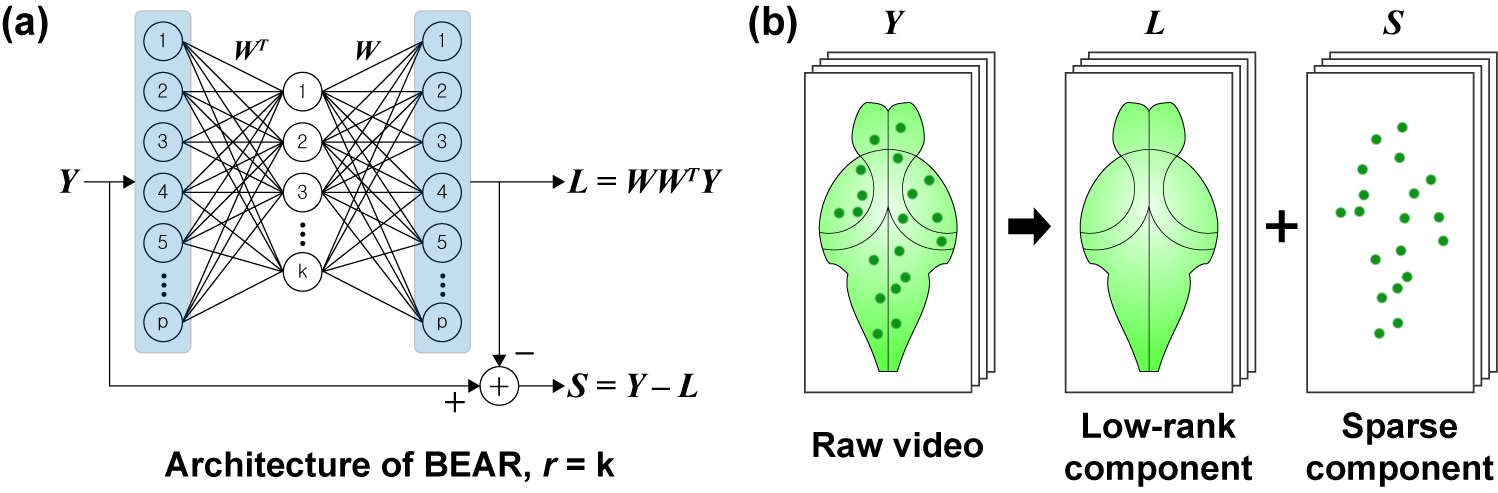}
\caption{Bilinear neural network for efficient approximation of RPCA (BEAR). (a) The low rank component $L$ is obtained by passing the input data $Y$ through the bilinear network. The sparse component $S$ is obtained as $S=Y-L$. (b) BEAR can decompose calcium imaging data into background and neural activity.}
\label{figure_1}
\end{figure}

BEAR is based on the following surrogate optimization:
\begin{equation}
\min_{L,S} ||S||_1 \textrm{ subject to } Y=L+S \textrm{ and } rank(L)\le r,
\end{equation}
which is attained by replacing the minimization of $||L||_{*}$ in (\ref{equation_2}) 
by the maximum rank constraint on $L$. 
A low-rank representation of the data is obtained by choosing a small integer $r$. 
The constraint $rank(L)\le r$ is enforced by setting $L=WMY$, where $W\in\mathbb{R}^{n\times r}$ and $M\in\mathbb{R}^{r\times n}$.
We note that $L=WMY$ is a stronger condition than $rank(L)\le r$, and 
this additional condition allows us to implement the optimization process as training the network shown in Fig.~\ref{figure_1}a.
Furthermore, we imposed an additional constraint on $W$ and $M$ by setting $M=W^T$ 
to reduce the number of trainable parameters and make the training more stable. 
Thus, the final form of the optimization problem is expressed as

\begin{equation}
\min_{W} ||S||_1 \textrm{ subject to } Y=L+S \textrm{ and } L=WW^TY.
\end{equation}

As the three constraints, (a) $rank(L)\le r$, (b) $Y=L+S$, and (c) $M=W^T$, are set by the network architecture,
the optimization problem can be solved by simply 
training the network to minimize the loss function $\mathcal{L}=||S||_1$ (Algorithm S1 in the supplementary material),
using gradient-based optimization algorithms, 
such as stochastic gradient descent, RMSprop~\cite{tieleman2012lecture}, and Adam~\cite{kingma2014adam}.
This allows us to use mini-batches for training the network;
hence, it can be applied to an extremely large data matrix $Y$.
This property is ideal for separating foregrounds in large-scale calcium imaging data (Fig.~\ref{figure_1}b). 
It should be noted that both forward propagation and backpropagation through the network require only two matrix multiplications thereby achieving the computational complexity of $O(nmr)$, whereas other methods that employ SVD have the computational complexity of $O(nm^2+n^2m)$.
Furthermore, the network can be used to infer the low-rank components without updating $W$, assuming that the network is trained with the earlier data points and the low-rank components remain stationary.
This inference-only mode can significantly improve the computation time.

\subsection{Extensions of BEAR} \label{2.2}

\subsubsection{BEAR with greedy rank estimation.}

BEAR requires the integer number $r$ to be explicitly set which has to come from prior knowledge regarding the rank of the low-rank matrix $L$, whereas general RPCA methods do not. To employ BEAR in the absence of such prior knowledge, we introduce an extension of BEAR with greedy rank estimation (Greedy BEAR).

Rank estimation is done by adding an outer loop to BEAR in which the target rank is gradually increased.
As soon as $rank(L)+\lambda||S||_1$ increases,
where $rank(L)$ is the target rank and $||S||_1$ is the loss after training the BEAR with the target rank,
the algorithm terminates,
and the network from the earlier iteration of the outer loop
is chosen as the final model.
Note that this algorithm (Algorithm S2 in the supplementary material) attempts to solve $\min_{L,S} (rank(L)+\lambda||S||_1)$, which is similar to the  original optimization problem expressed in (\ref{equation_1}).

\subsubsection{BEAR as non-negative matrix factorization.}
We note that BEAR can be modified to solve non-negative matrix factorization (NMF) by changing the loss function and introducing a non-negativity constraint on $W$ as follows:
\begin{equation}
\min_W ||R||_F \textrm{ subject to } Y = L + R \textrm{ and } L=WW^TY \textrm{ and } W\geq0.
\end{equation}
This optimization problem is equivalent to projective NMF~\cite{yuan2009projective}. It is known to produce localized sparse footprints, which is ideal for neuron segmentation. 

\subsubsection{BEAR cascaded with downstream networks.}

\begin{figure}[b!]
\includegraphics[width=\textwidth]{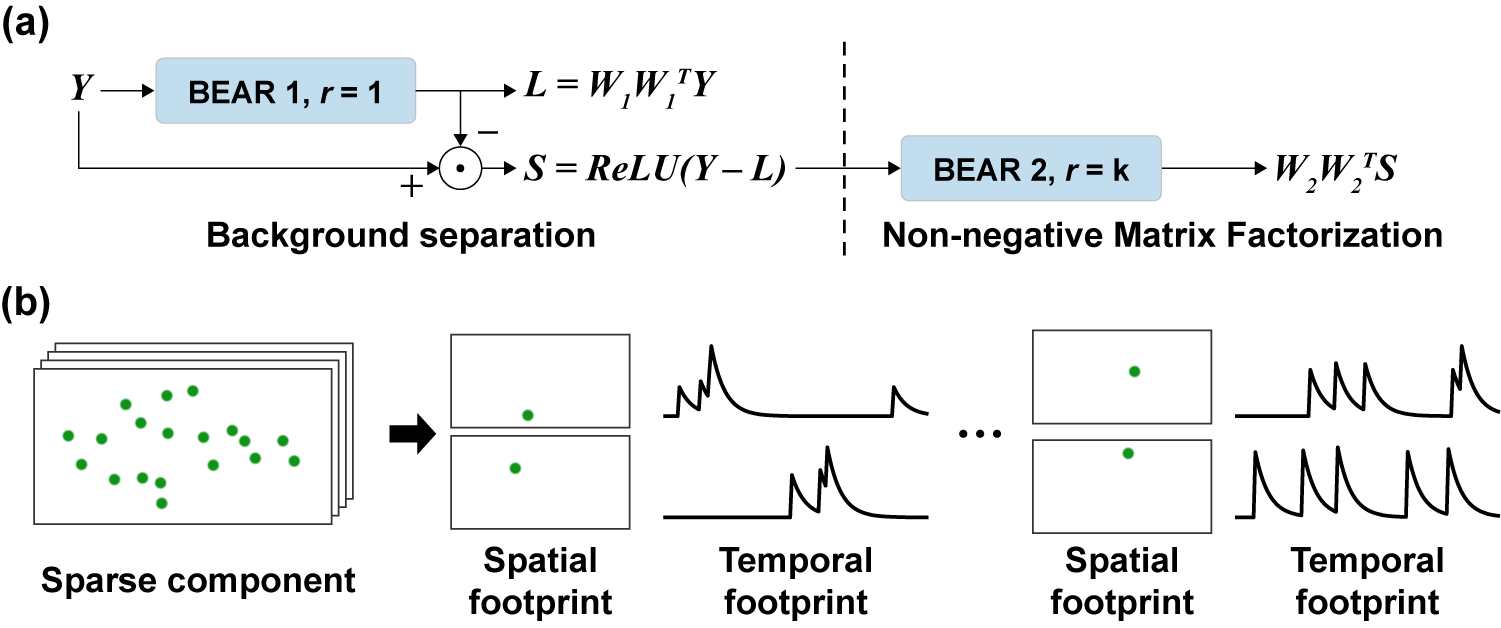}
\caption{(a) By cascading two BEARs, RPCA and NMF can be simultaneously performed. (b) The second BEAR in \textbf{a} takes sparse component $S$ from the first BEAR and performs NMF. The coefficient matrix $W_2$ and $W_2^TS$ correspond to the extracted spatial and temporal footprints, respectively.} \label{figure_2}
\end{figure}

Because BEAR is a neural network, it can be combined with other networks for downstream tasks and trained end-to-end.
This can be implemented by cascading BEAR with another network (Cascaded BEAR).
The loss function can be set as $\mathcal{L}=\mathcal{L}_1+\mu \mathcal{L}_2$,
where $\mathcal{L}_1$ and $\mathcal{L}_2$ are the loss functions
of the first and second networks, respectively,
and $\mu$ is a hyperparameter.
As an example, two BEARs can be cascaded
to simultaneously perform RPCA and NMF
as shown in Fig.~\ref{figure_2}a.
The sparse component from the first BEAR is fed to a ReLU layer
to ensure non-negativity and then fed to the second BEAR for NMF (Algorithm S3 in the supplementary material)
to extract the spatial and temporal footprints as illustrated in Fig.~\ref{figure_2}b.
\section{Experiments}

To validate and evaluate our algorithm,
we compared Greedy BEAR with existing RPCA algorithms, PCP~\cite{candes2011robust}, IALM~\cite{lin2010augmented}, GreGoDec~\cite{zhou2013greedy} and OMWRPCA~\cite{xiao2019online}.
Next, we applied the cascaded BEAR to calcium imaging datasets to demonstrate its capability to perform RPCA and NMF simultaneously.
We ran these tests on a PC with Intel i7-9700K CPU, NVIDIA GeForce RTX 2080 Ti GPU, and 128GB of RAM. The source code of PCP, IALM, and GreGoDec was from LRSLibrary~\cite{bouwmans2017decomposition,sobral2016lrslibrary}, a publicly available repository of multiple RPCA algorithms.
The source code of OMWRPCA was from~\cite{xiao2019online}.
PCP, IALM, GreGoDec, and OMWRPCA were run on the CPU.
BEAR was implemented using Pytorch,
and its performance on the GPU and the CPU was tested separately.
The computation time of BEAR includes the time taken for both training and inference.
The default maximum number of iterations of PCP was
reduced by a factor of 10.
The animal experiments conducted for this study were approved by the Institutional Animal Care and Use Committee (IACUC) of KAIST (KA2019‐13).

\subsection{Phase diagram in rank and sparsity}

\begin{figure}[t!]
\includegraphics[width=\textwidth]{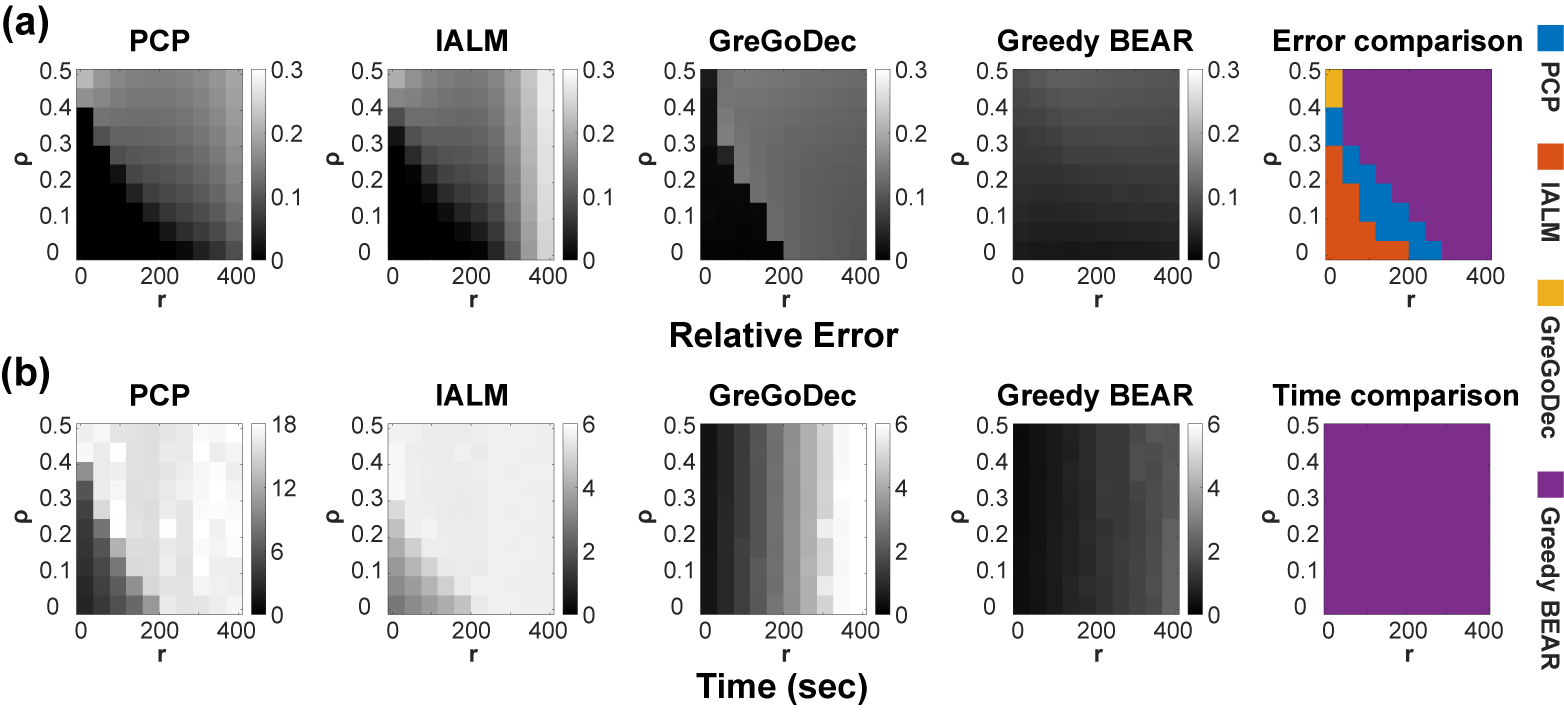}
\caption{Phase diagram for PCP, IALM, GreGoDec, and Greedy BEAR on $1000\times1000$ matrices.
(a) Relative errors.
(b) Computation times.
}
\label{figure_3}
\end{figure}

We evaluated the ability of the algorithms 
to recover varying rank matrices 
from data matrices that have error with varying levels of sparsity.
We randomly generated each data matrix $Y\in\mathbb{R}^{n\times n}$
by adding a low-rank matrix $L$ and a sparse matrix $S$,
where $n$ was set as 1,000.
Each low rank matrix was generated as $L=XY^T$,
where the entities of $X\in\mathbb{R}^{n\times r}$ and $Y\in\mathbb{R}^{n\times r}$ 
were drawn from a normal distribution
with the mean value of zero and the variance of $1/n$.
Each sparse matrix $S$ was generated 
by drawing each entity from a probability mass function as follows:
$P(x=0.1)=\rho /2, P(x=-0.1)=\rho /2, P(x=0)= 1-\rho$.

For each ($r$, $\rho$) pair, we generated 5 random matrices, 
each of which was solved with the four algorithms. 
The evaluation metrics were the relative error, defined as $||L-\hat{L}||_F/||L||_F$ where $\hat{L}$ was the recovered low-rank matrix, and the computation time. We averaged the relative error and computation time from 5 trials for each ($r$, $\rho$) pair.
We used the Adam optimizer with the batch size of $1,000$. The learning rate was $0.003$, and the number of training epochs was $50$.

Figure~\ref{figure_3} shows the obtained phase diagrams. Greedy BEAR achieved a relatively uniform level of error across a wide range of $r$ and $\rho$, while other algorithms failed in the high-rank or low-sparsity phase (Fig.~\ref{figure_3}a). Greedy BEAR was the fastest under all conditions (Fig.~\ref{figure_3}b).
In the rightmost plots in Fig.~\ref{figure_3},
each grid, which represents each ($r$, $\rho$) pair, is assigned to the best algorithm in terms of each measure. 
The percentages of ($r$, $\rho$) pairs which Greedy BEAR achieved the best relative error and the computation time were 69\% and 100\%, respectively.

\subsection{Performance on large calcium imaging data}

We measured the computation times of the algorithms using 3-D calcium imaging data. 
The data was acquired by performing calcium imaging of larval zebrafish brains at 4 Hz using a custom-designed epi-fluorescence microscope.
The larvae expressing GCaMP7a pan-neuronally were imaged at 4 days post fertilization. The  larvae were paralyzed in standard fish water containing $0.25 mg/ml$ of pancuronium bromide for 2 minutes prior to imaging
and then embedded in agarose for immobilization. 

\begin{table}[b!]
\caption{Computation times of four algorithms (in seconds). 
*Out of Memory. **Predicted based on small number of iterations. †Inference-only mode.}\label{tab1}
\centering
\begin{tabular}{c|c|c|c|c|c|c|c}
\hline
\multicolumn{1}{r|}{algorithms}            & PCP     & IALM   & GreGoDec & OMWRPCA     & \multicolumn{2}{c|}{Greedy BEAR} &  BEAR† \\ \cline{2-8} 
\multicolumn{1}{l|}{data size} & CPU & CPU & CPU & CPU & CPU & GPU & GPU     \\ \hline
$5313600 \times 150$ & 13814 & 1211 & 429 & 37883** & 371 & 134 & \textbf{42}    \\ \hline
$5313600 \times 1000$ & OOM* & OOM* & OOM* & 80883** & 1345 & 537 & \textbf{377}   \\ \hline
\end{tabular}%
\end{table}

\begin{figure}[t!]
\includegraphics[width=\textwidth]{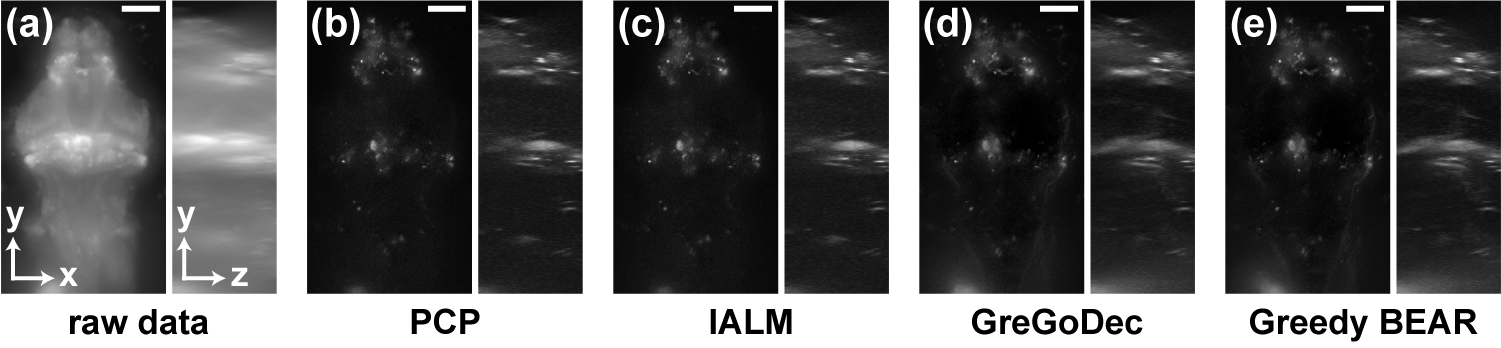}
\caption{Decomposition of the zebrafish calcium imaging data. Maximum-intensity projections along x and z axes are shown. (a) Input image.
(b-e) Sparse components from PCP, IALM, GreGoDec, and BEAR.
Scale bars, 100 µm.
} \label{figure_4}
\end{figure}

The sizes of the videos were $480(x)\times270(y)\times41(z)\times150(t)$ and $480(x)\times270(y)\times41(z)\times1000(t)$, respectively,
and they were reshaped to $5313600\times150$ (3.2GB) and $5313600\times1000$ (21.3GB), respectively.
For BEAR with inference-only mode, the network was trained with
the first one third of the data, and inference was performed on the entire data.
We used the Adam optimizer with the batch size of $64$. The learning rate was $0.00005$, and the number of training epochs was $45$.

The computation times of the four algorithms are summarized in Table~\ref{tab1}.
Greedy BEAR with GPU acceleration was faster than the existing algorithms and was capable of processing the largest data.
The decomposition results obtained by the four algorithms were visually nearly indistinguishable as shown in Fig.~\ref{figure_4} and Supplementary Video 1.

\subsection{Simultaneous RPCA and NMF using cascaded BEAR}

\begin{figure}[t!]
\includegraphics[width=\textwidth]{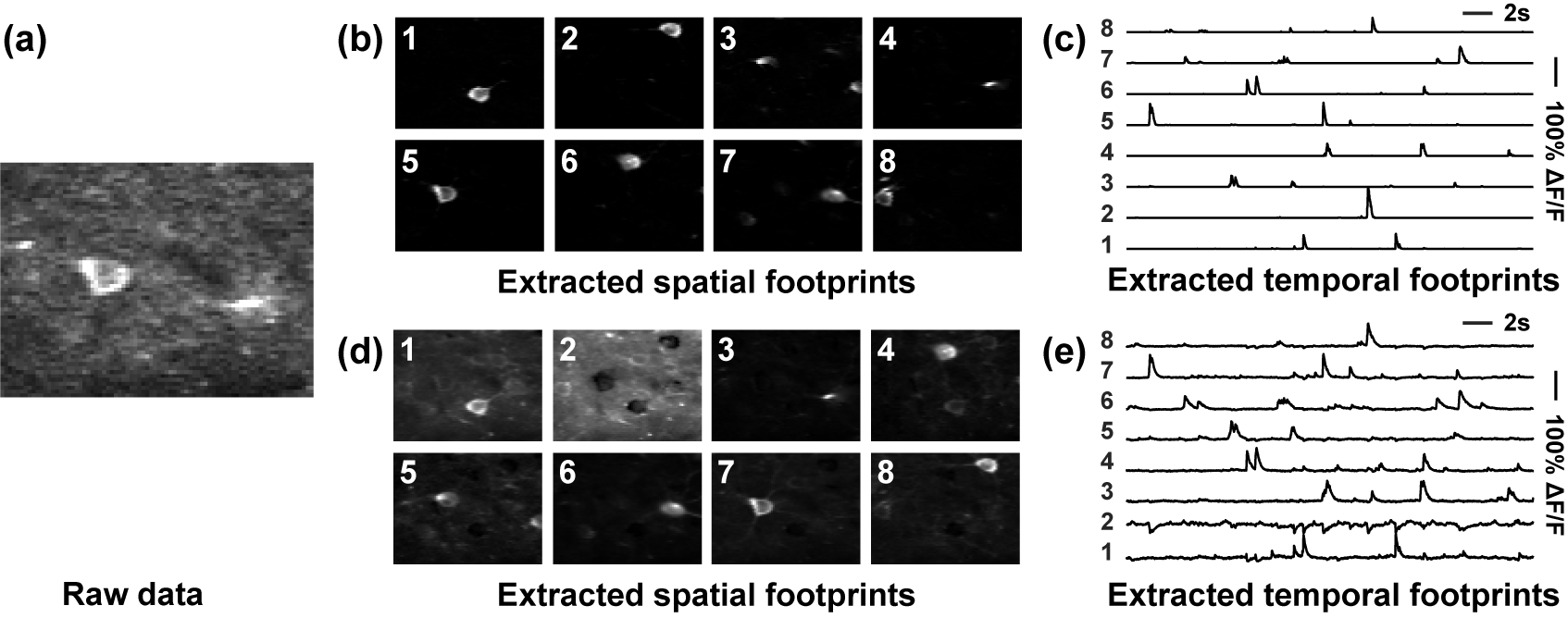}
\caption{Cascaded BEAR for analysis of neuronal activity of a mouse. 
(a) Single frame from the data. 
(b) Spatial footprints extracted using cascaded BEAR. (c) Temporal signals extracted using cascaded BEAR. 
(d) As in \textbf{b}, but using conventional NMF.
(e) As in \textbf{c}, but using conventional NMF.
}
\label{figure_5}
\end{figure}

We applied cascaded BEAR illustrated in Fig.~\ref{figure_2}a
to the publicly available mouse brain calcium imaging data obtained by two-photon microscopy~\cite{giovannucci2019caiman} for simultaneous RPCA and NMF. 
The dimensions of the video were $80(x)\times60(y)\times2000(t)$, which were reshaped to $4800\times2000$.
The ranks for the first and second BEARs were set as 1 and 8, respectively.
We used the Adam optimizer with the batch size of $512$. The learning rate was $0.0002$ and the number of training epochs was 5,000.
As shown in Fig.~\ref{figure_5}, the spatial components from cascaded BEAR were confined and sparse, 
whereas those from conventional NMF~\cite{berry2007algorithms,paatero1994positive} were not. In addition, the temporal signals from cascaded BEAR showed lower baseline fluctuation due to less signal mixing.

\begin{figure}[t!]
\includegraphics[width=\textwidth]{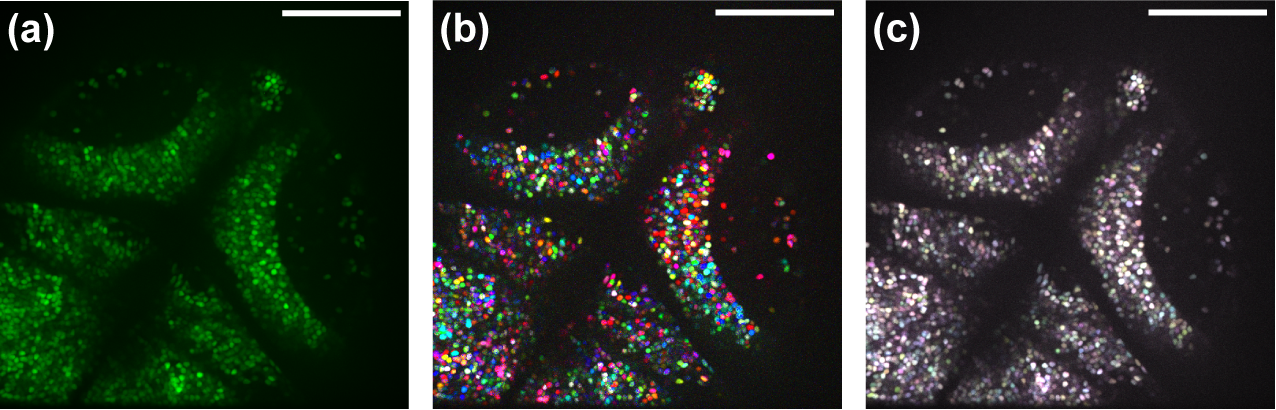}
\centering
\caption{Cascaded BEAR for analysis of neuronal activity of a larval zebrafish.
(a) A confocal image of a larval zebrafish brain expressing nuclear localized GCaMP6s.
(b) Spatial footprints from cascaded BEAR are colored and overlaid. 
(c) As in \textbf{b}, but from conventional non-negative matrix factorization.
Same random color palette as in \textbf{b} is assigned to each spatial footprint. Scale bars, 100 µm.
} \label{figure_6}
\end{figure}

Next, we applied cascaded BEAR to a zebrafish calcium imaging video (Fig.~\ref{figure_6}a).
It was acquired by imaging a larval zebrafish brain, expressing nuclear localized GCaMP6s pan-neuronally, at 10 Hz using a spinning disk confocal microscope with a 25x 1.1NA objective lens.
The dimensions of the video were $768(x)\times768(y)\times593(t)$, which were reshaped to $589824\times593$. The ranks for the first BEAR and second BEAR were set as 1 and 50, respectively. 
We used the Adam optimizer with the batch size of $64$. The learning rate was $0.0001$, and the number of training epochs was 1,000.
As shown in Fig.~\ref{figure_6}, cascaded BEAR was able to process the calcium imaging data with a very large number of neurons. 
The obtained spatial  footprints were confined and they corresponded well with the neurons (Fig.~\ref{figure_6}b), whereas the results obtained by conventional NMF were not confined, which manifested as low color contrast (Fig.~\ref{figure_6}c).

\section{Conclusion}

In summary, we proposed BEAR, a bilinear neural network for analyzing large calcium imaging data.
Because BEAR can be trained using the gradient from mini-batches,
it can decompose arbitrary sized data while exploiting GPU acceleration for speed improvement.
Furthermore, BEAR can be cascaded with other networks for end-to-end training.
For example, two BEARs can be cascaded for simultaneous RPCA and NMF to identify neurons in a large calcium imaging data.
Although our demonstration was focused on processing calcium imaging data, BEAR could be used for other general RPCA applications.

\subsubsection{Acknowledgements.}
This research was supported by National Research Foundation of Korea (2020R1C1C1009869), the Korea Medical Device Development Fund grant funded by the Korea government (202011B21-05), Institute of Information \& communications Technology Planning \& Evaluation (IITP) grant funded by the Korea government (MSIT) (No.2019-0-00075, Artificial Intelligence Graduate School Program (KAIST)),
and 2020 KAIST-funded AI Research Program. The zebrafish lines used for calcium imaging were provided by the Zebrafish Center for Disease Modeling (ZCDM), Korea.

\bibliography{references}
\bibliographystyle{splncs04}

\end{document}